\documentclass[sigplan,screen]{./acmart}
\pdfoutput=1
\usepackage[utf8]{inputenc}
\usepackage{microtype}
\usepackage{floatflt}
\usepackage{wrapfig}
\usepackage{listings}
\usepackage{booktabs}
\usepackage{color}
\usepackage{graphicx}
\usepackage{float}
\usepackage{amsmath}
\usepackage{tikz}
\usepackage{hyperref}
\usepackage{tablefootnote}
\usepackage{makecell}
\usepackage{colortbl}
\usepackage{marvosym}
\usepackage{blindtext}
\usepackage{longtable}
\usepackage{adjustbox}
\usepackage{pgfplots}
\usepackage{titlesec}
\usetikzlibrary{shapes.symbols,
  shapes.arrows, fit, patterns,
  arrows.meta, bending, shapes.callouts}
\usetikzlibrary{calc,positioning,shapes.geometric,shapes.symbols,shapes.misc}
\usetikzlibrary{arrows}
\usetikzlibrary{matrix}
\usetikzlibrary{arrows.meta}
\usetikzlibrary{graphs}
\usetikzlibrary{quotes}
\usetikzlibrary{datavisualization}
\usetikzlibrary{datavisualization.formats.functions}
\usetikzlibrary{positioning}
\usetikzlibrary{calc}
\usetikzlibrary{fit}
\usetikzlibrary{shapes}
\usetikzlibrary{tikzmark}
\usetikzlibrary{shadows.blur}
\usetikzlibrary{decorations.pathmorphing}
\usetikzlibrary{decorations.pathreplacing,calligraphy}
\newcommand\mpar[1]{\marginpar{\color{blue} \flushleft\sffamily\small #1}}
\renewcommand\mpar[1]{}

\newcommand\fig[1]{Fig.~\ref{fig:#1}}

\newcommand\secref[1]{Sec.~\ref{sec:#1}}

\newcommand{\project}[2]{\downarrow_{#1}(#2)}
\newcommand{\floor}[1]{\lfloor #1 \rfloor}

\newcommand{\twocolumns}[2]%
  {%
    \vspace*{0.03cm}
    \adjustbox{valign=t}{%
      \begin{minipage}[t]{\dimexpr(\linewidth)/2-\columnsep\relax}%
        #1
      \end{minipage}%
    }%
    \hfill%
    \adjustbox{valign=t}{%
      \begin{minipage}[t]{\dimexpr(\linewidth)/2-\columnsep\relax}%
        #2
      \end{minipage}%
    }%
  }%
\setcopyright{none}
\copyrightyear{}
\acmConference[IMPACT 2024]{14th International Workshop on Polyhedral Compilation Techniques}{January 17, 2024}{Munich, Germany}
\acmDOI{}
\acmISBN{}
\acmYear{}
\acmPrice{}

\begin{document}

\title[%
  Employing polyhedral methods to optimize stencils on FPGAs
]{%
  Employing polyhedral methods to optimize stencils on FPGAs
  with stencil-specific caches, data reuse, and wide data bursts%
}

\author{Florian Mayer, Julian Brandner, and Michael Philippsen}
\email{{florian.andrefranc.mayer,julian.brandner,michael.philippsen}@fau.de}
\affiliation{
  \institution{Programming Systems Group,
    Friedrich-Alexander Universität Erlangen-Nürnberg (FAU), Germany}
  \country{}
}
\begin{abstract}

    It is well known that to accelerate
    stencil codes on CPUs or GPUs and to exploit
    hardware caches and their lines optimizers must find spatial and
    temporal locality of array accesses to harvest data-reuse
    opportunities. On FPGAs there is the burden that there are
    no built-in caches (or only pre-built hardware descriptions
    for cache blocks that are inefficient for stencil codes).
    But this paper demonstrates that this lack is also a
    chance as polyhedral methods can be used to generate
    stencil-specific cache-structures of the right sizes
    on the FPGA and to fill and flush them efficiently
    with wide bursts during stencil execution. The paper
    shows how to derive the appropriate directives and
    code restructurings from stencil codes so that
    the FPGA compiler generates fast stencil hardware.
    Switching on our optimization improves the runtime of a set of 10 stencils
    by between $43\times$ and $156\times$.

\end{abstract}

\maketitle
\begin{tikzpicture}[remember picture, overlay]
    \node[xshift=1cm, yshift=-0.5cm] at (current page.north west)
    [text width=193.9mm, inner sep=1mm, 
    fill=red!20, rounded corners, anchor=north west, align=left]
    {
  Published at the 14th International Workshop on Polyhedral Compilation Techniques, (IMPACT 2024, in conjunction with HiPEAC 2024), Munich,
Germany, Oct. 17, 2024, 12 pages, see
\url{https://impact-workshop.org/impact2024/#mayer24-fpgas} \\
A replication package and the benchmarks can be found at:
\url{https://dx.doi.org/10.5281/zenodo.10396084}
    };
\end{tikzpicture}

\section{Introduction}\label{sec:intro}

In mature polyhedral compilers, loop tiling is
a common optimization when targeting CPUs or GPUs.
It restructures a loop
such that it processes the iteration
space block-by-block and it constructs tiles that access most of the
array elements often and in the same order in which they
reside in memory, i.e., with temporal and spatial locality.

There have been attempts to offload codes to FPGAs, i.e., to
use a hardware synthesis (HLS) to generate circuits for them on the
FPGA. However, out of the box tiled loops \cite{isl1,polyAstGen}
in general do not turn into efficient FPGA circuits.
There are two main obstacles. First, the HLS cannot turn the tiled
loop into hardware pipelines (as the loop boundaries and increments
are too complex for the HLS to detect patterns). Without hardware
pipelining the benefits of FPGA accelerators vanish.
Second, on CPUs or GPUs tiled loops run faster since their data locality
exploits the built-in cache hierarchy. As on FPGAs there are no
caches, there is no benefit from tiling alone. In general,
the HLS
generates from the tiled loop a hardware block (kernel) with
connections to an FPGA bus through which the kernel directly
talks to the DDR, i.e., without any host-side cache support.
There are two approaches to still allow
loop tiles to enjoy data locality.
One can either add an FPGA hardware
block that implements a cache between the kernel and the DDR. FPGA
hardware vendors offer such pre-fabricated hardware descriptions.
Such cache circuits on the FPGA have been used to accelerate FPGA
kernels \cite{paperCache}. Similar to host-side cache hardware
they only come with a few different sizes and they only can
load lines of adjacent data
and predict what to best replace. But if cache circuits
need to be added to the FPGA anyway one can do better than using a
generic cache block.
For stencil codes this paper shows that by using
polyhedral methods to understand the access patterns
we can statically determine (a) the stencil-specific best cache sizes
and (b) pre-load exactly what is needed
and spill what is no longer needed.
But the polyhedral method can also (c) help improve the burst
behavior.
While off-the-shelf caches have a fixed line size on the side to the
DDR and a fixed word size on the side to the running code, a better
memory bandwidth can be reached with wide (and stencil-specific)
bursts. We show how to use polyhedral methods to achieve the data
alignment requirements.
Even more, instead of asking the HLS to construct such
a stencil-specific cache and add it between
kernel and DDR, we (d) inline the cache functionality, i.e.,
we  use buffer arrays and load operations within the stencil.

The main contribution of this work is a polyhedral-based code
generator for FPGAs. It uses loop tiling to improve locality, adds
buffers and directives to generate inlined cache-like hardware
circuits that exploit that locality, and calculates a data-shipment
that uses wide bursts to fill these caches.

\secref{not}
introduces our loop tiling terminology.  \secref{caches} covers our
cache buffers, their types, and their fusion. Both sections set
the stage for explaining our optimization in \secref{impl}.  \secref{rw}
reviews related work before we present our quantitative results in
\secref{eval} and conclude.

\begin{figure}[tb]%
  \renewcommand{\baselinestretch}{0.75}\footnotesize%
\begin{lstlisting}
// Original loop
for (i=1; i<N; i+=1)
  A[i] = B[i-1] + B[i+1];

// Tiling 1: with $SZ=32$ (here $\Delta=0$)
// iterations per tile: 32, 32*, rest
for (ti=1; ti<N; ti+=32)
  for (i=max(1, ti); i<min(N, ti+32); i+=1)
    A[i] = B[i-1] + B[i+1];

// Tiling 2: with $SZ=32$ and $\Delta=1$
// iterations per tile: 32-1=31, 32*, rest
for (ti=1; ti<N; ti+=32)
  for (i=max(1+1,ti); i<min(N-1, ti+32); i+=1)
    A[i-1] = B[i-1-1] + B[i+1-1];

// Tiling 3: with $SZ=32$ (normal form, $\delta=(0,0)$)
// iterations per tile: 31, 32*, rest
for (ti=0; ti<N; ti+=32)
  for (i=max(1, ti); i<=min(N-1, ti+31); i+=1)
    A[i] = B[i-1] + B[i+1];

// Tiling 4: with $SZ=32$ and $\delta=(-1,0)$
// iterations per tile: 32, 32*, rest
for (ti=1; ti<N; ti+=32)
  for (i=ti; i<=min(N-1, ti+31); i+=1)
    A[i] = B[i-1] + B[i+1];

// Tiling 5: with $SZ=32$ and $\delta=(-1,-ti)$
// iterations per tile: 32, 32*, rest
for (ti=1; ti<N; ti+=32)
  for (i=0; i<=min(31, N-ti-1); i+=1)
    A[i+ti] = B[i-1+ti] + B[i+1+ti];

// Tiling 6: with $SZ=32$ and $\delta=(-1,-ti)$, padded
// iterations per tile: 32, 32*, rest
for (ti=1; ti<N; ti+=32)
  for (i=0; i<=31; i+=1)
    A[i+ti] = B[i-1+ti] + B[i+1+ti];
\end{lstlisting}%
\caption{Loop Tiling Example.}\label{fig:minEx}%
\end{figure}%

\section{Notation for Tiling Transformations}\label{sec:not}
An optimizing compiler tiles a loop by cutting its iteration space
into segments of size $SZ$, so-called tiles. The result is an outer
loop (inter-tile loop, index $ti$) that holds an inner loop
(intra-tile loop, original index $i$, original body).  In
Fig.~\ref{fig:minEx} we tile the given loop with a tile size of
$SZ=32$. Since the original iteration space may not be divisible by
$SZ$ the first and/or the last tiles in general have fewer than $SZ$
iterations.  On CPUs/GPUs the purpose of tiling is to enhance cache
locality. When the code accesses a memory address, the built-in cache
hardware loads a full line that also includes adjacent addresses. In
the ideal case, the tile size fits to the size of the cache lines and
when a tile loop accesses an address in its first iteration, the cache
hardware pre-loads all the data that later iterations of the tile need
for faster accesses without cache-misses. However, this effect is only
visible if the boundaries of the tiles are properly aligned w.r.t.\
the addressing requirements of the cache hardware. Hence, in addition
to picking a tile size, the optimizing compiler also picks a $\Delta$
that (slightly) shifts the boundaries of the tile loops, i.e., the
first tile starts $\Delta$ iterations later and is shorter by $\Delta$
iterations.  The inner tiles keep their sizes.  The last tile grows by
$\Delta$ iterations.  In Fig.~\ref{fig:minEx} the Tiling 1 does not
have such a shift ($\Delta = 0$). The second one uses $\Delta =
1$. Such a shifting of the tile boundaries is important when targeting
an architecture with built-in cache hardware. However, when targeting
an FPGA, it is necessary to generalize the concept of shifting to both
the outer and the inner loop. This turns $\Delta$ into a vector
$\delta = (\delta^o,\delta^i)$. For a simpler formalism, we also
calibrate the outer loop to start from 0 as shown in Tiling 3. We call
this the normal form, $\delta=(0,0)$. Unfortunately, the HLS cannot
generate an efficient FPGA from the normal form. Only if its analysis
can prove that the loop boundaries are runtime constants, the HLS
unrolls a loop into parallel/pipelined circuits.
The \texttt{max}
and \texttt{min} operations in the boundaries of the inner tile loop
hide from the HLS that this is the case.
By shifting the outer loop by
$\delta^o=-1$ its first index value matches the starting index of the
original loop (and this also ensures that the first tile is a full
tile of size $SZ$). The more important effect is that the \texttt{max}
operation is no longer needed, as Tiling 4 shows. But the HLS still
does not understand the lower bound expression \texttt{i=ti} of the
inner loop. Only by shifting the inner loop by $\delta^i=-ti$ so that
it starts from a constant \texttt{0}, we can make the HLS generate
unrolled and parallel circuits for it. Note, that because of the
normalization for any vector $\delta$, i.e., for the last three
tilings, the following predicate holds for the loop indices:\\
\centerline{$(ti + \delta^o)\ mod\ SZ = 0 \land ti \leq i < ti + SZ$}

In addition, some compilers also pick an execution order $o$ for the
iterations of a tile, e.g., to run a loop backwards. Formally, the
bijective ordering function $o$ maps an index $i$ of the tile loop to
an $o(i)$. The lexicographic order of the values $o(i)$ is the
execution order of the tile. While for any choices of $SZ$ and
$\delta$ a tiling is always valid, altering the order $o$ may break
the data dependences that exist in the original loop. This paper does
not take reordering into account.

Loop tiling can recursively be applied to each of the loops in a given
loop nest of depth $d$. To achieve the intended effects on locality of
accesses, compilers in general first re-group the resulting $2d$ loops
and interchange them so that all the $d$ inter-tile loops surround
all the $d$ intra-tile loops. They then pick a permutation $p$ of size
$d$ that shuffles/interchanges the $d$ inner (tile) loops to select
the tiling with best locality. Re-grouping and interchanging
the loops in a loop nest may break data dependences that exists in
the original loop nest.

The take-away is, that when ignoring iteration space reorderings
and taking the initial re-grouping into inter-tile
loops followed by intra-tile loops as given, a \textit{tiling
transformation} $T$ for a $d$-dimensional loop nest is fully specified
by a triple $(SZ, p, \delta)$ with a vector of tile
sizes SZ=(SZ$_1$, \dots, SZ$_d$), a permutation $p$
indicating how to shuffle the intra-tile loops, and a delta vector
$\delta=(\delta_1^o, \delta_1^i, \dots, \delta_d^o, \delta_d^i)$.
Formally, for the indices of
the resulting $2d$ loops the effect of $T$ is:
\begin{equation*}
\begin{split}
  T(SZ, &p, \delta) = \{ O[i_1, ..., i_d] \to \\
        &\qquad O[ti_1, ..., ti_d, p(i_1) +
          \delta_{p(i_1)}^i, ..., p(i_d) + \delta_{p(i_d)}^i] : \\
          & \bigwedge\nolimits_{x=1}^d
   (ti_x + \delta_x^o)\ mod\ SZ_x = 0 \land ti_x \leq i_x < ti_x + SZ_x \}
\end{split}
\end{equation*}

Note, that the lexicographical order $O[i_1, ..., i_d]$ in which the
$d$ original loops are executed is replaced by a $2d$-di\-men\-sion\-al
order for the resulting loop nest of depth $2d$.

\begin{figure}%
  \renewcommand{\baselinestretch}{0.75}\footnotesize%
\begin{lstlisting}
for (i=1; i<N-1, i++)
  for (j=1; j<N-1; j++)
    for (k=1; k<N-1; k++)
S:    V[i,k,j] = V[i,k,j] + A[i,j,k]
               + A[i+1,j+1,k+1];
\end{lstlisting}%
\underline{SCoP of the above loop nest:}\hfill\phantom{*}%
\begin{align*}%
  Dom &= [N] \to \{ S[i, j, k] : \qquad\qquad\qquad\qquad\qquad\qquad\qquad\qquad \\
        & 1 \leq i < N - 1 \land 1 \leq j < N - 1 \land 1 \leq k < N - 1 \} \\
Sch &= \{ S[i, j, k] \to O[i, j, k] \} \\
R_0 &= \{ S[i, j, k] \to V[i, k, j] \} \\
R_1 &= \{ S[i, j, k] \to A[i, j, k] \} \\
R_2 &= \{ S[i, j, k] \to A[i+1, j+1, k+1] \} \\
W_0 &= \{ S[i, j, k] \to V[i, k, j] \}
\end{align*}%
\caption{Running Example.}\label{fig:origStencil}%
\end{figure}

The polyhedral toolbox \cite{poly,poly3,poly4,poly2,poly1,scop1} is a
set of operators and functions that help work with tiling
transformations (and other loop optimizations). To do so, the toolbox
relies on some formalism describing loop nests, the data accesses in
their bodies, and the index sets involved (today often encoded
with the Integer Set Library (ISL)
\cite{isl1}). The key element of the formalism is a so-called Static
Control Part (SCoP) that represents a loop nest.  It consists of the
domain set $Dom$ that models the iteration space of the loop nest, a
schedule $Sch$ that defines the order of iterations by mapping
iterations to a lexicographic order $O$, and a list of access
relations for reading $R$ and writing $W$. They represent the array
accesses of the individual iterations. A loop nest can be represented
with a SCoP if in its code all conditions and loop bounds are affine
functions of runtime constants and loop indices. In this paper we
require all array accesses to also be in that form. \fig{origStencil}
holds a $d=3$ dimensional running example and its SCoP. The schedule
$Sch$ maps the iteration vector $(i,j,k)$ to the lexicographic
order. When handed this SCoP, the polyhedral toolbox can detect that
$R_0$ and $W_0$ address the same data element in every iteration $(i,
j, k)$. \fig{tiledStencil} is the result of applying a tiling
transformation. In the resulting SCoP, only $Sch$ changes. $Sch'$ turns
the execution order into a 6-dimensional lexicographic one and it
restricts the index values as defined by $T(SZ,p,\delta)$ above.

\begin{figure}%
\renewcommand{\baselinestretch}{0.75}\footnotesize%
\begin{lstlisting}
// outer loops, inter-tile
for (ti=0; ti < N-1; ti += $SZ_i$)
  for (tj=0; ...)
    for (tk=0; ...)
      // inner loops, intra-tile, p=(i,k,j)
      for (i=max(...); i<=min(...); i++)
        for (k=max(...); ...)
          for (j=max(...); ...)
            S(i, j, k);
\end{lstlisting}%
\underline{Modified parts of the SCoP:}\hfill\phantom{*}%
\begin{align*}%
  Sch' = &\{ S[i, j, k] \to O[ti, tj, tk, i, k, j] :
           \quad \qquad\qquad\qquad\qquad\qquad \\
  & ti\ mod\ SZ_i = 0 \land ti \leq i < ti + SZ_i \land \dots \\
  & tk\ mod\ SZ_k = 0 \land tk \leq k < tk + SZ_k \}
\end{align*}%
\caption{Normal form of tiling the code in Fig. \ref{fig:origStencil}
with the transformation
  $T(SZ=(SZ_i, SZ_j, SZ_k), p=(i,k,j), \delta=(0,0,0,0,0,0))$.
  S is a shorthand of the original loop body.%
}\label{fig:tiledStencil}%
\end{figure}%

\section{Cache buffers: Types and Fusion}\label{sec:caches}
\textbf{\textit{Cache Buffers for Working Sets of Tiles.}}  Assume a
stencil had only one single array access. The fundamental idea of our
approach is to use a stencil-specific cache buffer that holds all the
data elements that a single tile needs, i.e., that -- for a fixed set
of loop indices of the $d$ outer loops -- all iterations of the $d$
inner loops of a tiling access. We call these data elements the
\textit{working set} of a tile. Such a cache can improve the total
runtime of a tile because of two reasons: First, it may be possible to
apply an efficient burst load to fill the cache (spatial locality)
before the tile starts its execution and to then serve the tile with
faster accesses to the cached data instead of accesses to the copies
residing in the DDR and, second, to benefit even more if the tile
accesses cached data elements more than once (temporal locality).  For
general $n$-point stencils we aim to use one buffer per array access.

For the single access to a $d$-di\-men\-sional array \texttt{A}, a
tile with sizes $SZ=(SZ_1,\dots,SZ_d)$ has a working set of
$\prod_{1}^{d} SZ_i$ data elements (for many types of stencils) that
can be stored in a $d$-di\-men\-sional buffer array of size
$\prod_{1}^{d} SZ_i$. We call this a \textit{full} buffer (of
\texttt{A} in a tiling). If the running example only had the access to
\texttt{A[i+1,j+1,k+1]} then a buffer array of size $32^3$ could be
filled before the start of a tile's execution so that the tile could
work with the data in that buffer. Whereas for the stencil of the
running example, the tile sizes suffice to compute the size of the
buffer array, in general, it is necessary to calculate the minimal
\textit{bounding box} around the data elements in the working
set.\footnote{From the working set $ws = A(f)$ for an access
  relation $A$ and a set of iterations of a full tile $f = [ti_1,
    \dots, ti_d] \to \{ O[ti_1, ..., ti_d, i_{p(i_1)}, \dots,
    i_{p(i_d)}] \}$ we project out the $d$ dimensions $dims =
  proj(ws)$ by means of $proj = \{ [i_1, \dots, i_d] \to Ai_1[i_1];
  \dots; [i_0, \dots, i_d] \to Ai_d[i_d] \}$. The corners of the
  bounding box are the lexicographically smallest ($lmins=lmin(dims)$) or largest ($lmax(dims)$) index values per
  dimension.
  To calculate the extent of the bounding box, we build a set per
  dimension that contains all elements between the smallest and the
  largest element. Thus, for each universe set $u$ in
  $S(univ(dims))$, where S is the space decomposition operator, we
  build the one-dimensional interval set $in$ by $lb(lmin) \cap
  ub(lmax)$, where $lb = (lmins \cap u) \preccurlyeq univ(lmins \cap
  u)$ and $ub = (lmaxs \cap u) \succcurlyeq univ(lmaxs \cap u)$.  The
  intersection of the inverse projections of the $d$ interval sets,
  $\bigcap_{i=1}^{d} proj^{-1}(in_i)$, is the bounding box of $ws$. }
For example, if instead of \texttt{j+1} the second index is
\texttt{2*j}, we would need a buffer twice as large to hold all the accessed array elements (plus the unused ones between them).
Only
if the size of the bounding box is statically known, a buffer array
can be used for caching. Otherwise we label the array access as
\textbf{nc} (no caching).

There are ways to use smaller caches/buffers without losing
the advantages of pre-loading. Instead of pre-loading the full working
set of a tile, we look at the sub-tiles
that the outermost intra-tile loop
processes. Such a sub-tile has the same $d$-dimensional array access
in its body and hence, in general, it needs the same full buffer to
store its working set. But for a fixed index of the outermost tile
loop, the bounding boxes of the array accesses often are much smaller
than the tile sizes. In this case, we pre-load only those elements
into the smaller cache (ideally with a burst load) before a sub-tile
starts. We call such a buffer a \textit{chunk} buffer.  In the running
example for a fixed value of \texttt{i} (the index of the
outermost tile loop), the bounding box of the working set has a width
of 1 in the $i$-dimension. Thus all the elements that the sub-tile
accesses only need a much smaller buffer of size 1$\cdot$32$\cdot$32.

Chunk buffers need fewer FPGA resources than full buffers, leave
more FPGA floor space for the tiles' circuits, and allow
larger tiles with more parallelism. But there is a
downside. As long as there is a pre-loading phase followed by a tile's
execution phase and a write-back phase that is needed in case of write
accesses, in general there is a runtime penalty due to redundant data
loadings. Consider a chunk of data that the outermost tile loop loads
into the chunk buffer. Whenever the loop ticks, most of the data
elements in the buffer are likely to be loaded again, albeit into a
separate slice of the buffer array. While the chunk buffer is smaller
than the full buffer, the runtime performance in general drops due to
the redundant loadings. The chunk buffer is only a wise choice
for tiling permutations that allow to shift the data locally
within the buffer instead of reloading it from the DDR. Then a chunk
buffer can be used even more efficiently by partly overlapping the
pre-loading and the sub-tile's execution. Instead of pre-loading all
its elements, it is possible to pre-load only the data elements that
the first few iterations of the sub-tile need. And while those
iterations are busy, the FPGA-circuits \textit{concurrently} pre-load
only what the next iteration of the sub-tile will need. Whenever a
sub-tile iteration is done, the data in the chunk buffer shifts along
its outermost axis.  Note that it depends on the tiling permutations
whether a chunk buffer is useful. It can be applied if each sub-tile
needs a data chunk of the same size and if the data shift between any
two adjacent sub-tiles only depends on the index of the outermost tile
loop. For the permutation $p = (i,k,j)$ of the running example we
cannot use a chunk buffer for the access \texttt{A[i+1][j+1][k+1]} as
$p$ does not permit data bursts to fill it. Only for $p = (., ., k)$
we can pick a chunk buffer for \texttt{A}.

For the chunk buffer we kept the index of the outermost intra-tile
loop fixed. A generalization is to fix the index values of all but the
innermost intra-tile loops, i.e., to only look at its working set,
that, as before, often can be stored in a smaller cache buffer.  There
is a special case with an even smaller FPGA footprint and more
concurrent pre-loading: If for the array access the sizes of the
working set bounding boxes are 1 in all but one of its dimensions,
then a one-dimensional array can be used as a so-called \textit{line}
buffer.  In the running example, assume fixed values for \texttt{i}
and \texttt{k}, then for \texttt{A[i+1][j+1][k+1]} the size of the
one-dimensional buffer array is $1 \cdot 32 \cdot 1$.  Similar to the
situation with chunk buffers, conceptually all tiling permutations
could use a line buffer, but only some of them allow for a pre-loading
with data bursts while others require a much slower initialization of
the complete line.  For the running example we cannot pick a line
buffer, as the access to \texttt{A} is not in line w.r.t. the inner
\texttt{j} loop (and thus slow).

\paragraph{Fusion of Caches for Temporal Locality.}
Now that we have discussed how to achieve spatial locality with
smaller cache buffers and how the best choice depends on the tiling
permutation, let us improve the general idea to $n$-point stencils that have
more than one array access and that may
have data dependences among their accesses. We aim to exploit the data
reuse potential and save on the number and total sizes of the cache buffers
needed for such stencils.

If a stencil has two accesses to the same array (both without
\textbf{nc}-label), there are two cases that matter: (a) One can
use a combined cache buffer for both of them
if the size of the combined cache buffer is smaller than the sum of
the sizes of the two individual cache buffers. (b) It may not be
possible to use a cache buffer for any of the accesses if there is a
(loop carried) data dependence.

To check if a case applies, we use the polyhedral toolbox to compute
the intersection of the working sets of the two array
accesses.\footnote{Consider one of the array accesses, say
  \texttt{A[i,j,k]} which is $R_1$ in the SCoP.
  To compute its working set, the term
  $R_1'$=$(R_1^{-1}\ . \ Sch')^{-1}$ turns the original access relation $R_1$
  in the given SCoP into $R_1'$ that reflects the schedule $Sch'$
  after tiling and maps $O[ti_1,\dots,ti_d,i_1,\dots,i_d] \to
  A[i_1,\dots,i_d]$. Once we have pre-processed the two access relations that
  way, we can apply the set of all iterations in a full tile
  ($full = [ti_1, ..., ti_d] \to \{ O[ti_1, ..., ti_d, i_{p(i_1)},
    ..., i_{p(i_d)}] \}$) to both of them and compute the intersection
    $R_1'(full) \cap R_2'(full)$. (Same for $W$ access relations.)
  }%
For an \textit{empty}
intersection (not shown in the running example)
we use separate cache buffers and pick their types as discussed above.
For a \textit{non-empty} intersection there are
two cases. If there is no data dependence between the two accesses, we
\emph{can} use a shared cache, fuse their working sets,
and pick the buffer type that fits the union of the sets.  We only
fuse if this reduces the total memory demands and if it does not turn
the fused accesses into \textbf{nc} ones. For example, there may be
array accesses that, individually, have working sets suitable for
small chunk (or even line) buffers. But the union of
their working sets may need a full buffer or -- even worse -- may not
be suitable for caching at all due to a statically unknown size.  If
there is a data dependence, we either \emph{must} use a shared cache for both
working sets or refrain from using caches at all.  Separate caches may break the data
dependence, as a value written into one cache may not be observed by a
read from the other cache. We label both array accesses
as \textbf{nc} and do not use a cache buffer for them at all.  In
general, there can be a set of accesses (not just two) that form a
graph of data dependences. We label all those accesses as \textbf{nc}
as soon as there is an empty intersection of the working sets of any
pair of accesses in that set. (This is a simplification: there may be
some pairs in the set that could use a shared cache buffer without
breaking the data dependence. This paper ignores the optimization potential.)

\newcommand{\attrSep}[0]{0.41ex}%
\tikzset{%
  qfix/.style={minimum height=3.5em}%
}
\begin{figure}%
\centering%
\small%
\begin{tikzpicture}[%
    ogn/.style={
      draw, rectangle,
      thick, align=center,
      minimum width=9em
    },
    attr/.style={
      inner sep=0ex,
      align=center,
      minimum width=9em,
    },
    column 1/.style={anchor=center},
    column 2/.style={anchor=center},
    column 3/.style={anchor=center}
    ]
\node[matrix] (mat) {%
      &
      \node[attr] {\underline{$p = (i,k,j)$}}; &
      \node[attr] {\underline{$p = (i,j,k)$}}; \\
      \node[ogn] (f0) {$A[i,j,k]$,\\$A[i+1,j+1,k+1]$}; &
      \node[attr, qfix] {full buffer \\ \texttt{A'=[33,33,33]}}; &
      \node[attr, qfix] {chunk buffer \\ \texttt{A'=[2,33,33]}}; \\
      \node[ogn] (f1) {$V[i,k,j]$,\\$V[i,k,j]$}; &
      \node[attr, qfix] {chunk buffer \\ \texttt{V'=[1,32,32]}}; &
      \node[attr, qfix] {full buffer \\ \texttt{V'=[32,32,32]}}; \\
      \node[inner sep=0ex, right] { total cost: }; &
      \node[attr] {33$^3$+1$\cdot$32$^2$ = 36\ 961}; &
      \node[attr] {2$\cdot$33$^2$+32$^3$ = 34\ 946}; \\
    };
\end{tikzpicture}
\caption{Fused working sets and selected cache buffers
  (types and sizes) for two out of
  six feasible permutations for the running example with
  tile sizes $SZ=(32,32,32)$.
}\label{fig:graphs}
\end{figure}%

In the running example the two accesses to \texttt{A} have a non-empty
intersection. The same holds for both accesses to \texttt{V}.  On its
left side, \fig{graphs} shows the fused accesses. We apply the
selection process discussed above to pick the suitable cache buffer
types for them and to gauge their sizes (and thus implicitly their
potential for concurrent pre-loading).  Recall that this choice
depends on the permutation.  If the shared buffer does not save space
we keep the buffers separate.

The two working sets of two array accesses of \texttt{A} are
different from the union of those sets.  The tile accesses in each of
the array dimensions an element plus its adjacent element. The sizes
of the bounding boxes of the combined working set are thus
$[33,33,33]$ instead of $[32,32,32]$ for both of the individual
accesses.  Let us first consider the permutation $p=(i,k,j)$ that the
running example uses. Here the decision process described above cannot
pick chunk buffers (as $p$'s last index isn't $k$) and thus picks
\textit{two} full buffers of size $32^3$ for the individual
accesses. For the combined working set it picks \textit{one} full
buffer \texttt{A'} of size $33^3$. Since this is smaller, we fuse and
use a shared buffer, see the middle column of \fig{graphs}.

If the running example had used a different permutation $p=(i,j,k)$
there would be chunk buffers for both individual \texttt{A} accesses,
each with size 1$\cdot$32$\cdot$32.  The size of the bounding box of
the $i$-dimension of the combined working set is 2 because the tile
accesses both \texttt{A[i][.][.]} and \texttt{A[i+1][.][.]}.
\textit{One} chunk buffer \texttt{A'} of size 2$\cdot$33$\cdot$33 is
better than \textit{two} chunk buffers of size
1$\cdot$32$\cdot$32.
In its last column, \fig{graphs} shows this
choice. (Rationale: The innermost loop index \texttt{k} allows fast
pre-loading from \texttt{A'}, the working sets of each sub-tile for
fixed \texttt{i} all have the same sizes, and their data chunks can be
shifted with a constant offset to avoid re-reads.)  At first
glance, the savings do not seem to be large. However, as discussed
before, the data can now quickly be shifted locally within the fused
buffer instead of reloading the same data slowly from the DDR (per
tick of the outermost tile loop).

For the two accesses to \texttt{V} in the running example the union of
their working sets is identical to the individual working sets. In
case of the permutation $p=(i,k,j)$ we save by using one full buffer
for it. For $p=(i,j,k)$ even a chunk buffer can be used, see
\fig{graphs} for the sizes.

A comparison of the two permutation columns of \fig{graphs} reveals
that their total costs, i.e., the sums of the sizes of the necessary
cache buffers are different.  Selecting the cheaper permutation
$p=(i,j,k)$ not only leaves more FPGA floor space for larger tile sizes
and hence more parallelism. But smaller buffers also allow for faster
and even concurrent pre-loading, without losing their ability to
exploit data reuse.

\section{Optimization Method}\label{sec:impl}

Our optimization can process $n$-point stencils that live in a ca\-no\-nical
loop nest with a rectangular iteration domain, that can be represented
by a SCoP, and that work on non-overlapping $m$-dimensional rectangular
arrays with arbitrarily many reads or writes.  Except for the array
accesses there may not be statements with other observable side effects
(although we allow for floating point divisions by zero). Our
optimization works on in-place stencils, but excludes those where
tiling breaks the data dependences of the stencil (e.g. Gauss-Seidel
or SOR methods).

Our optimization works in five steps. Step~1 picks the
tiling sizes $SZ$. Step~2 selects a permutation $p$ of the inner tile
loops and also assigns possibly shared cache buffers to the accesses.
Step~3 chooses the deltas $\delta$ that shift tile boundaries
for better HLS generated hardware. Step~4 redirects the indexing expressions 
from the original array to instead address the
assigned cache buffers. Step~5 generates loop code, inserts
buffer declarations, suitable HLS directives, and data shipment.
Below we discuss these steps in detail and illustrate them with the
running example in \fig{origStencil}.

\paragraph{Step 1 -- Picking the tile sizes SZ}

Whereas for CPUs or GPUs with built-in cache hardware, the tile sizes
need to fit the fixed line size of the cache, conceptually we can pick
any tile sizes when offloading stencils plus stencil-specific cache
functionality to an FPGA. As larger tiles in general cause more
parallelism by unrolling plus pipelining and thus run faster, we aim
at the largest possible tile sizes that the FPGA floor can
hold. However, since larger tiles need larger caches that compete for
the floor space, the combined resource consumption limits the possible
tile sizes. In an ideal world, an autotuner \cite{autotune} can find
the best configuration.

In the real world there is another limit. It is the time that
the high-level synthesis takes to generate the specification of the
FPGA circuits. This time strongly grows with the tile sizes (and the
parallelism that they cause). We found that when allowing 3-5 hours to
perform the HLS for a stencil code in our benchmark set
(see \secref{eval}), the largest practical tile sizes for our
benchmarks are $SZ$=$(1024)$ for $d$=$1$ stencils, $SZ$=$(128,128)$ for
$d$=$2$, and $SZ$=$(32,32,32)$ for $d$=$3$ codes.

\paragraph{Step 2 -- Picking a tiling permutation
with the smallest cache buffers}

We generate the set of all feasible permutations $p$ of the inner tile
loops and purge those that violate data dependences. For each
remaining permutation, we assign caches to the array accesses as described
in \secref{caches}, i.e., we pick a buffer type for an array access or
decide that it cannot be cached (either because the size of its
boundary box is not statically known or because there are data
dependences that forbid caching), and whenever possible, we fuse
accesses to save space with their shared buffers. We then calculate
the total size of all assigned buffers and
pick the permutation with the lowest cache demand (smallest total
cost). When there is a tie, we pick the permutation that uses faster
buffers (\emph{line} faster than \emph{chunk}, \emph{chunk} faster
than \emph{full}).

For the running example all 6 feasible permutations are valid as the
stencil does not have any loop carried dependences. \secref{caches}
does not assign the \textbf{nc}-flag to any of the accesses. For two
of the permutations we discussed the
total cost above. As the other 4 permutations are not cheaper, we
pick $(i,j,k)$ and the buffer assignment shown in \fig{graphs}.

\paragraph{Step 3 -- Picking a delta and an iteration padding}%
\label{sec:step3}

Consider \fig{minEx} again. When the HLS generates FPGA circuits
from the code of Tiling 3, there is no concurrency -- the FPGA
processes the iterations sequentially.  The HLS cannot unroll/pipeline
the tile loop in a way that builds concurrent circuits for its
iterations. Current HLS tools can only unroll/pipeline loops whose
bodies have statically known sizes (when unrolled recursively). In
particular the obstacles are that the
\texttt{max}- and the \texttt{min}-expressions used for the
bounds of the inner tile loop are too complex for the pattern matching
of the HLS, and that the tile loop starts from a symbolic \texttt{ti}
that the HLS does not identify as being runtime constant. Step 3 picks
a delta vector that gets rid of these obstacles.

\fig{minEx} shows in Tiling 4
the effect of $\delta$=$(-1,.)$.
By setting $\delta^o$ to the negated lower bound of the original
untiled loop we get rid of the \texttt{max}-operation, extend the
first tile to a full tile, and have an inner tile loop that starts
from \texttt{i=ti}. We apply this $\delta^o$-shift to all loops of a
given SCoP by using polyhedral operations\footnote{ $Sch' := Sch\ .\ Adj$,
where $Adj = tra(\Delta(tra(lmin(Sch(Dom))))^{-1})$. Note that $Sch$
and $Dom$ are of the ISL type union set and union map, resp. For the
sake of simplicity we omit the space decomposition operator here.
$tra$ is ISL's translate operation.
}
that affect the schedule $Sch$ of the loop nest, see the $\delta^o$
terms in the $T(SZ,p,\delta)$-formula in \secref{not}.

The delta $\delta$=$(.,-ti)$ in \fig{minEx} causes the inner tile loop
to start from a constant \texttt{0} (Tiling 5).  By setting
$\delta^i$=$-ti$ we get rid of the symbolic \texttt{ti} and shift the
inner tile loop by an offset of \texttt{ti} to then start
from \texttt{i=0}.  We also apply this $\delta^i$-shift to all loops
of a given SCoP by using polyhedral operations\footnote{ $Sch' :=
Sch\ .\ Zs$, where $Zs = \{ O[ti_1, ..., ti_d, i_1, ..., i_d] \to O[ti_1,
..., ti_d, p(i_1) - \delta^i_{p(ti_1)}, ..., p(i_d)
- \delta^i_{p(ti_d)}] \}$.} that again affect the schedule $Sch$, see
the $\delta^i$ terms in the $T(SZ,p,\delta)$-formula in \secref{not}.

At this point there is only the \texttt{min}-operation left that keeps
the HLS from generating unrolled parallel circuits for the innermost
tile loop. There are two approaches to get rid of this. One option is
to use index set splitting on the outer loop so that (a) all the full
tiles have the same fixed number of iterations and can hence be
unrolled, while (b) the incomplete last tile is cut off and processed
separately. The consequences are not only a higher consumption of FPGA
resources due to the duplication, but also a slower runtime because
the separate incomplete tile is unrolled less often which limits the
overall throughput. We therefore pursue the other option: we round up
the number of iterations in the last tile to the nearest integer
multiple of the tile size, i.e., we pad the iteration space of the
innermost tile loop, see Tiling 6 in \fig{minEx}. Note, that the extra
iterations use hardware circuits that are already present, as full
tiles also need them. The extra iterations do not cost runtime as the
hardware circuits process them concurrently to the other iterations of
the last tile. We implement the padding by modifying the domain $Dom$
of the SCoP with polyhedral operations that take the lower and upper
bounds plus the stride of the loop as well as the tile size into
account and compute new upper bounds and a $mod$-operation that
ensures that the resulting loop has the same stride as
before.\footnote{For the SCoP's $Dom$ let $S =
stride(\project{0}{Dom})$ denote the stride of the innermost loop
$i_d$ and let $lb(Dom) = lmin(\project{0}{Dom})$ be the lower bound of
this loop.  Then $segStarts = \{ S[i_1, ..., i_d] \to S[i_1,
..., \floor{\frac{i_d - lb(Dom)}{SZ_d}} SZ_d + lb(Dom)] \}$ is the
stream of index values at which the tiles start, i.e., the values of
$ti$. From these starts we add $SZ_d$ indices for each of the tiles,
in particular for the last one.  $padding = \{ S[i_1, \dots, i_{d-1},
i_d] \to S[i_1, \dots, i_{d-1}, \pi] : i_d \leq \pi < i_d + SZ_d \land
(\pi - i_d)\ mod\ S = 0\}$.  In the resulting $Dom' :=
padding(segStarts(Dom))$ the $mod$-operations implement the stride of
the original loop. For simplicity we omitted the permutation
$p$ from all the loop indices of the tile
loops.}  In the resulting Tiling 6 the inner tile loop always iterates
32 times, even if \texttt{i} reaches or surpasses \texttt{N}. As the
buffers that are needed for the full tiles are sufficiently large to
also hold the data that the padded last tile needs, there is no need
for larger buffers.

Obviously we need countermeasures to prevent that the padded
iterations change the semantics of the loop by writing to previously unused
slots of the cache buffers or by working with uninitialized slots.

Let us talk about the two cases of write accesses first.  (a) If the
write access is not cached (\textbf{nc}), for example if it plays a
role in a data dependence, we wrap it in a guarding \texttt{if}
statement that keeps the extra iterations from writing.  The branch
statement hampers instruction level parallelism in the generated
hardware, but as the HLS generates specific hardware, the negative
effects are less severe than for a CPU's micro-instruction pipeline.
(b) If the write access is cached, we trim the write back from the
cache to the DDR accordingly, i.e., values in the cache buffer that
are a result of padded iterations are not written back and hence no
not affect the semantics of the loop. There is a corner case worth
mentioning: assume the stencil would compute a value that is based on
the number of iterations. It must store that value somewhere and
modify it in every iteration. Hence, there is a write-after-write
dependences which causes the above \textbf{nc}-case and thus the
guards that switch off the effects of the added iterations. Hence,
corner cases like this work correctly.

Whereas on CPUs computations with uninitialized values, e.g., a
division by zero, may trigger exceptions that may give away the
presence of padded iterations, FPGAs switch to NaN instead and thus
hide the presence of the padding.

\paragraph{Step 4 -- Index redirection}\label{sec:indexRedir}

Whenever we have assigned a buffer cache to an array access, the
original array access needs to be retargeted to address the cache
buffer instead of the original array. This is straightforward in case
of full buffers, as it suffices to replace the name of the original
array with the name of the cache buffer. All the index expressions in
the code can remain unchanged. For chunk and line buffers however, we
need to redirect the original index expressions.  The new index
expressions depend on the loop indices of the inter- and intra-tile
loops, which in turn depend on the tiling transformation
$T(SZ,p,\delta)$), and also on an offset relative to the upper left
corner of the cache buffer. Let us skip the discussion of the lengthy
and complex symbolic computations on the SCoP that make use of the
polyhedral toolbox. Instead we refer both to a detailed discussion of
a similar task in a PPCG-paper \cite{ppcg} and to the online data set
and replication package \cite{replPack} that accompanies this article.

To illustrate the index redirections at work, we consider the
permutation $(i,j,k)$ in \fig{graphs}.  Instead of \texttt{V} we just
access the full buffer \texttt{V'} which can be done with a simple text
replacement in the SCoP of the loop nest. To retarget the
stencil's two accesses to \texttt{A} to instead access the
chunk buffer \texttt{A'}, we both replace \texttt{A[i,j,k]}
with \texttt{A'[0,j,k]} and \texttt{A[i+1,j+1,k+1]} with
\texttt{A'[1,i+1,j+1]}. Here the first index
expression of both redirected accesses no longer depends on the loop
index \texttt{i} because the chunk buffer slice shifts by 1 whenever
the $i$-loop ticks.

\paragraph{Step 5 -- Code generation with declarations,
  pragmas, data shipment, halos, and bursts}\label{sec:dataShipment}
\begin{figure}%
\renewcommand{\baselinestretch}{0.75}\footnotesize
\begin{lstlisting}
// outer loops, inter-tile
for (ti=1;...) for(tj=1;...) for (tk=1;...) {
  float V'[32, 32, 32]; float A'[2, 33, 33];
  #pragma HLS ARRAY_PARTITION V' complete
  #pragma HLS ARRAY_PARTITION A' complete

  //fill V':
  ship(V,[ti,tj,tk],V',[0,0,0],3,1);
  // fill b1-1 2d-slices of A':
  ship(A,[ti,tj,tk],A',[0,0,0],2,1);
  for (i=0; i<=min(...); i++) {
    // fill b2-1 rows of A':
    ship(A,[ti+i+1,tj,tk],A',[1,0, 0],1,1);
    for (j=0; j<=min(...); j++) {
      #pragma HLS PIPELINE
      // fill next row of A':
      ship(A,[ti+i+1,tj+j+1,tk],A',[1,j+1,0],1,1);
      for (k=0; k<32; k++) // padded loop
        V'[i,k,j] = V'[i,k,j] + A'[0,j,k]
                  + A'[1,j+1,k+1];
    }
    // shift 1 slice of A' by 1:
    ship(A',[1,0,0],A',[0,0,0],2,1);
  }
  //flush V':
  ship(V',[0,0,0],V,[ti,tj,tk],3,1);
}
\end{lstlisting}
\caption{Generated code for the running example in \fig{origStencil}
  for $T$($SZ$=$(32,32,32)$, $p$=$(i,j,k)$,
  $\delta$=$(-1,-ti_1,\dots,-1,-ti_d))$
  with shared buffers \texttt{V'}=[32,32,32],
  \texttt{A'}=[2,33,33] from \fig{graphs}.}\label{fig:transformed}
\end{figure}

\newcommand{\ship}[0]{\texttt{ship}}

Here we generate code from the SCoP.  For each of the buffer array, we
add a static declaration just before the inner loops (inter-tile). The
code in \fig{transformed} shows this for the running example.  For each
buffer we also add a pragma
\texttt{ARRAY\_PARTITION} that tells the HLS to map the array to
FPGA registers \cite{xilinxhls}. This prepares them for the
concurrent accesses in the inner tile loop that the HLS
unrolls/pipelines because of the added pragma
\texttt{PIPELINE}. Note, that the HLS automatically duplicates parts of
the buffer arrays to resolve access conflicts
during pipelining. Hence, we do not need to add double buffering.
But earlier works \cite{alias} manually
added it to work around older HLS.

\paragraph{Data shipment.}
Depending on the type of buffer, we use different strategies to
fill it before it is accessed. We initialize
a \textit{full} buffer completely before the first iteration of the
tile and fully flush it after its last iteration.
In \fig{transformed} the first \ship{} fills the buffer \texttt{V'} of
the running example completely, i.e., it reads 1 (= last argument)  single
3-dimensional (= next-to-last arg.) block of \texttt{V} starting
from \texttt{V[ti,tj,tk]}. This reads what the upcoming tile
$(ti,tj,tk)$ needs. Let us postpone the discussion of the
(pseudo-) macro \ship{}.

In their outermost dimensions the size a \textit{chunk} array depends
on the width of the bounding box of the index values that a tile
computes.  If an $i$-loop accesses [i][]\dots[] and [i+$c$]\dots[],
the width is $c$+1 and we pre-load $c$ consecutive $d$-1-dimensional
data segments into the buffer before the tile starts. One level down
in the intra-tile loops, the array's bounding box $b_2$ may be larger than
the tile size $SZ_2$.  We then pre-load $b_2$-$SZ_2$ slices of
$d$-2-dimensional data.  In general, the number of segments that
a \ship{} loads is the difference between the size of the bounding box
and the tile size in the corresponding dimension.

In \fig{transformed}, there are three \ship{} operations that fill the
chunk buffer \texttt{A'} before the tile's body runs.  The
first \ship{} loads one (= last arg.) 2-dimensional chunk (=
next-to-last arg.) of data to the buffer, as the bounding box has size
$c$=2 in the most significant dimension.  Afterwards one 2-dimensional
slice with sizes $[1,33,33]$ remains to be loaded.  Since in the next
dimension down the intra-tile loops, we have $b_2$=33 and $SZ_2=32$, the
second \ship{} fills a line of \texttt{A'}.  Both the line
with the $i$-loop and the index in the source array \texttt{A} tick.
Just before the innermost loop we \ship{} the remaining line
into \texttt{A'}. Its indices into \texttt{A} move with both the $i$-
and the $j$-loop. After the three data shipments the chunk buffer is
complete and the tile body can run. Then a \ship{} shifts the
data within the chunk buffer so that some data (needed by later
sub-tiles) remain in the buffer, albeit at a different index; the
shift also makes room to (pre-)load new data for the next sub-tile.

\begin{figure}%
\renewcommand{\baselinestretch}{0.75}\footnotesize%
\begin{lstlisting}
// src, src-offset, dest, dest-offset
// $di$ dimensional segment, $R$ repetitions
ship(S,  [$os_{d}, \dots, os_{di+1}, os_{di}, \dots, os_{1}$],
     D, [$od_{d}, \dots, od_{di+1}, od_{di}, \dots, od_{1}$], $di$,  $R$);

for ($r$=0; $r$<$R$; $r$++)
  for ($s_{di}$=0; $s_{di}$ < $DS_{di}$; $s_{di}$++)
    ...
      for ($s_2$=0; $s_2$ < $DS_2$; $s_2$++)
        burstcpy(D[$od_{d}, ..., od_{di+1}, od_{di}+s_{di}+ci, ...,
          od_{1}+s_1$],
          S[$os_{d},..., os_{di+1}, os_{di}+s_{di}+ci,
          ..., os_{1}+s_1$],$DS_1$);
\end{lstlisting}%
\caption{The semantics of the ship macro.}\label{fig:ship}%
\end{figure}%

\paragraph{Halos and bursts.} The \ship{}
(pseudo-) macro expands into a nest of loops, one per array dimension,
see \fig{ship}.  The additional enclosing loop deals with the number
of repetitions if, as in case of the initial \ship{} of the chunk
buffer \texttt{A'}, a few segments need to be copied. The innermost
array dimension is copied in \texttt{memcpy}-style (instead of another
loop that would copy values one element at a time).

If we actually used \texttt{memcpy}, the HLS would generate standard
DMA operations that, in general, process one array value per cycle of
the FPGA hardware. Luckily, by specifying a so-called \emph{port
width} $w$ we can direct the HLS to generates a channel of width $w$
that handles $w$ values per cycle. This bursts the filling and
flushing of buffers by a factor of $w$ and significantly improves
performance, see \secref{eval}.  However, there is a downside: Bursts
require that both the DDR-side memory address and the number of array
elements to be shipped are is divisible by $w$.

To exploit the bursting ability we thus need to extend buffers with
slightly larger ones, more specifically, we round up the size of the
innermost array dimension to the next size that is divisible by $w$.
In the example, for $w$=4 we actually declare and allocate an
array \texttt{A''} of size [2,33,\textbf{36}] that is large enough to
host the buffer \texttt{A'} with halo elements around it. But how many
of the 3 halo elements in the last dimension live on which side of the
buffer \texttt{A'}? To answer this question we need to look at the
first element of the original array \texttt{A} that needs to be copied
into the buffer. If this array element is not aligned w.r.t. to the
$w$-addressing requirements, we need to copy a few more values left of
it. The number of extra values determines how many halo
elements \texttt{A''} needs to have on the left of the
buffer \texttt{A'}.\footnote{In our implementation, step 2 of course
works with the sizes of the halo-added buffers. And the index
redirections in step 4 take the number of halo elements into account
and compute replacement indices that are displaced by the number of
needed halo elements. And finally, the code generator earlier in step
5 declares buffer arrays with halos and uses ship{} operations that
use the displaced indices. We hid the index displacements so far as
they would have affected the digestibility of the explanations through
the paper.}

\section{Related Work}\label{sec:rw}
Our work is related to a range of different publications.  Many works
use polyhedral methods for reuse detection in a software environment.
Many approaches \cite{poly4,Kodukula,wolf,meister1,meister2}
reorder memory accesses
to benefit from the multilayer-multipurpose caches present on modern
CPUs and some \cite{meister1,meister2} also employ explicit
multi-buffering to scratch-pads,
but these concepts cannot easily be transferred to standard FPGA
architectures that are not equipped with such caches or scratch-pads.
PPCG
\cite{ppcg} is a source-to-source compiler that uses the polyhedral
model to generate SIMD parallelism for a GPU.  Despite generating CUDA
code instead of HLS C the approach is similar to ours as it uses
temporary arrays to buffer reused data.  It does not address aspects
like bursts or pipeline performance, as these are unique to hardware
design.  The method is evaluated on ten benchmarks from the PolyBench
collection, including the 6 stencil codes that we also look at. They
do not evaluate on the Adept benchmark from our set.
Dtg\_codegen \cite{dtgCodegen} generates SIMD instructions
for machine learning codes. Similar to our step 3, it
adds padding iterations when the iteration space cannot
be tiled evenly, but it does not address burst or pipeline
performance.
The model from
Van Achteren et al. \cite{vanAchteren}
calculates from loop nests how to best exploit scratch-pad
memories, but it does not employ multiple buffer types, does not
optimize the data shipment, and does not use loop tiling.

Some authors
\cite{paperCache,inlineCaches,betterInlineCaches,multipurposecustom}
add multipurpose caches to FPGA architectures to exploit data reuse
and to speed up performance.  This approach typically does not require
any complex analysis of the input code and therefore does not utilize
polyhedral methods.  In turn, the heuristics of multipurpose caches
limit the performance gain and also severely limit the possibilities
for bus widening.  However, our technique is limited to homogeneous
and statically determined reuse patterns, whereas multipurpose caches
have no such limitation.  Since we can detect array accesses that we
cannot cache (\textbf{nc}-label), combining our approach with a
multipurpose cache for such accesses may be beneficial.  All four
works ignore established benchmark sets and evaluate only on small,
proprietary sets of eight or fewer benchmarks.

Liu et al. \cite{loopSpl} utilize polyhedral methods in an FPGA
environment.  Their approach uses polyhedral analysis to perform loop
splitting and thereby increases the loop pipelining performance.  In
contrast, our goal is to increase the effective storage throughput,
and we therefore generate vastly different hardware.
They only present quantitative results for one stencil code (2d jacobi)
and
achieve small speedups (2.7) compared to
our 17 -- 39 on the same 2d jacobi kernel, probably
because they do not use any buffers.
On the other hand, their approach is not limited to stencil code
and could therefore be used to speed up loops where our method
is inapplicable.

Friebel et al. \cite{FiebelTensor} translate fluid dynamic tensors
from a DSL to HLS C. They use the polyhedral model to find
dependences and reschedule operations to reduce their distances.
This optimization is potentially orthogonal to ours, but
does not employ any reuse or burst optimizations.

Closer related to us are works that generate FIFO channels to buffer
reusable data \cite{Cong2016,Meeus,Natale,autosa}. These works, like
ours, use polyhedral methods in the FPGA domain to improve on data
reuse.  However, their approach to hardware generation is
significantly different from ours.  Instead of building cache
structures, they use the polyhedral method to split the stencil code
into multiple hardware blocks and connect them via FIFO channels
through which reusable data is sent to the next consumer. The method
promises a smaller resource utilization than ours but cannot be fully
implemented on HLS source level, and therefore cannot utilize modern,
highly optimized HLS tools. It also does not allow for any burst
optimizations and bus widening.
Since in contrast to us, the AutoSA research tool \cite{autosa}
can handle non-canonical loop nests, we cannot evaluate their
benchmark codes in our systems. While it would be interesting
to study and compare the FPGAs that AutoSA generates from our
stencil codes, we could not upgrade their system from the
older Vitis HLS 2019.2 that they used to the current one.
The numbers in the papers cannot be compared because
of the different optimizations that the HLS versions provide.
Meeus et al. \cite{Meeus} and Natale et al. \cite{Natale}
evaluate only about
half the stencils we look at.
They mainly focus on resource or energy efficiency
instead of runtime performance.
Cong et al. \cite{Cong2016} and Wang et al. \cite{autosa} do not show
any results on stencil codes.

Issenin et al. \cite{drdu} use polyhedral methods to improve
data-reuse on ASICs by employing scratch-pad RAM. As the
work is not focused on a specific problem area, it has to rely on
highly complex models that on larger codes require extreme runtimes.
(In practice, the user can cut the optimization short but has to
accept sub-optimal results.)  The publication therefore evaluates the
performance with only four algorithm snippets, all from the field of
computer graphics. We use a much broader set of stencils in our
evaluation.

Pouchet et al. \cite{pbdr} propose a framework for optimizing stencil
codes. One of its features is a polyhedral-based data-reuse
optimization similar to ours. However, they only cover cases where
reuse occurs between two consecutive iterations of the innermost
stencil loop.
Our work has none of these limitations.
Their method is inapplicable to about half of our benchmarks set.
For most of the remaining stencils
their restrictions lead to many more memory accesses than our method
(3x for 2D-5P).

Deeus et al. \cite{deest} and Alias et al. \cite{alias} both
present FPGA code generators that use tiling to
exploit intra-tile \emph{and} inter-tile
data reuse.
While our approach can
be orthogonally extended with their
inter-tile reuse method,
our intra-tile reuse method is more advanced as it
defines three buffer types, picks them depending on
the tiling permutation, and is unique in using
wide bursts to increase the bandwidth.

\section{Evaluation}\label{sec:eval}

To evaluate we use all 4 stencil codes from the Adept benchmark suite
\cite{adept} and 4 (out of 6) stencil codes from PolyBench
\cite{polybench}.  We exclude \texttt{adi} as it uses a non-canonical
loop nest. The data dependences in \texttt{seidel} prevent loop
tiling.  Since in \texttt{fdtd} there are three distinct sub-stencils
that we evaluate separately, we have a set of 10
codes. Table~\ref{benchmarks} holds some relevant characteristics. The
set has a range from simple 4-point stencils up to a 27-point
stencil. Note that there is a lot of data-reuse. Step 2 of
\secref{impl} can always fuse the accesses into 2 or 3 shared
buffers. Except for the \texttt{fdtd}-benchmarks that update their
arrays in place, all codes use shadow arrays to compute new values and
hence do not have any (loop-independent) dependences.

{%
\renewcommand{\baselinestretch}{0.9}\small
\begin{table}
  \caption{Characteristics of the Benchmark Set.}
  \begin{tabular}{lcrccr}
    name & dimen-& stencil & shared & in- & $F_{max}$ \\
    & sions & points  & buffers & place & in MHz \\
    \hline
1D-jacobi & 1 &  4 & 2 & no  & 124 \\
2D-5p     & 2 &  5 & 2 & no  & 261 \\
2D-9p     & 2 &  9 & 2 & no  & 205 \\
2D-jacobi & 2 &  6 & 2 & no  & 200 \\
2D-fdtd0  & 2 &  4 & 2 & yes & 345 \\
2D-fdtd1  & 2 &  4 & 2 & yes & 320 \\
2D-fdtd2  & 2 &  6 & 3 & yes & 260 \\
3D-19p    & 3 & 19 & 2 & no  & 145 \\
3D-27p    & 3 & 27 & 2 & no  &  55 \\
3D-heat   & 3 & 11 & 2 & no  &  50 \\
  \end{tabular}
\label{benchmarks}
\end{table}
}

We use the OpenMP-to-FPGA compiler ORKA-HPC \cite{ORKA} to offload
code fragments to the FPGA that are flagged with the \texttt{target}
pragma.  Our Ubuntu 20.04.4 LTS test system with an Intel Core i7-4770
CPU is connected to a Xilinx VCU118 FPGA board via PCI express (width
x4, 5GT/s). We synthesize the FPGA hardware with Xilinx Vitis 2021.2.
TaPaSCo DSE \cite{TAPASCO} finds the highest possible frequencies
($F_{max}$ in Table~\ref{benchmarks}) for the hardware synthesis of
the codes.  But to ease comparison, we always used 50 MHz to measure
the speedups (even though higher frequencies would yield higher
speedups).  To derive SCoPs from the canonical loop nests of the
stencils we rely on the Polyhedral Extraction Tool (PET) \cite{pet}.

For each benchmark, we measure and average the runtimes of 5 runs,
excluding the time it takes to initially ship the data from the host
to the DDR memory on the FPGA board. Below we discuss the speedups
that our optimized FPGAs achieve in comparison to the baseline, i.e.,
the FPGA that the HLS generates with its default settings. The
baseline does not use any buffering (and there is no
burst-enhanced data shipment), but there is an implicit
\texttt{PIPELINE} pragma in the innermost stencil loops. We use fixed
seeds to generate the same random input arrays (with 64 MiB of data)
for all measurements. All FPGA
versions have the same results as a sequential run of the
stencil on the CPU.

\newcommand{\labelPos}[0]{(0.5,-0.52)}
\begin{figure}
\pgfplotsset{width=\linewidth, height=4cm}
\centering
\begin{tikzpicture}
\begin{axis}[
    ybar=1.44pt,
    bar width=5pt,
    ylabel={speedup},
    symbolic x coords={
        1D-jacobi,2D-5p,2D-9p,2D-jacobi,2D-fdtd0,
        2D-fdtd1,2D-fdtd2,3D-19p,3D-27p,3D-heat},
    xtick=data,
    x tick label style={rotate=40, anchor=east, inner sep=0pt},
]

\addplot[pattern=crosshatch dots gray] coordinates {(1D-jacobi, 79.59616893)(2D-5p,
50.47070317)(2D-9p, 87.67462558)(2D-jacobi, 59.93383129)(2D-fdtd0,
74.50505991)(2D-fdtd1, 64.90257981)(2D-fdtd2, 36.78994654)(3D-19p,
42.08542198)(3D-27p, 47.81842197)(3D-heat, 21.23561143)};
\addplot[pattern=north east lines] coordinates {(1D-jacobi, 109.7340615)(2D-5p,
70.06574608)(2D-9p, 134.1088598)(2D-jacobi, 86.26127252)(2D-fdtd0,
156.9009508)(2D-fdtd1, 0)(2D-fdtd2, 76.33073038)(3D-19p,
96.83004876)(3D-27p, 117.7512451)(3D-heat, 43.90319504)}
node[pos=0.4, anchor=west, xshift=.8ex] {\small156}
node[pos=0.95, anchor=west, xshift=.8ex] {\small43}
;

\end{axis}
\end{tikzpicture}
\caption{ Speedups of two optimized FPGAs for varying tile sizes (and
  port widths 16).  Small Tiles
  \protect\tikz\protect\node[draw,rectangle,pattern=crosshatch dots
  gray, minimum width=1em] {};: $SZ_{1D}$=$(512)$, $SZ_{2D}$=$(64,64)$,
  $SZ_{3D}$=$(16,16,16)$.  Large Tiles
  \protect\tikz\protect\node[draw,rectangle,pattern=north east lines,
  minimum width=1em]
  {};: $SZ_{1D}$=$(1024)$,
  $SZ_{2D}$=$(128,128)$, $SZ_{3D}$=$(32,32,32)$.  Empty bar: HLS failed with
  a timing error.}\label{fig:tiles}
\end{figure}

\paragraph{Impact of the tile sizes.} For all 10 benchmarks (ordered by
data dimensionality), \fig{tiles} shows the speedups over the baseline
for large tile sizes with 3-5 hours of HLS time. Halving the sizes in
each dimension (= small tiles) causes the speedup to drop
significantly, by a factor of $1.3$$\times$ to $2.5$$\times$.  Larger tiles are
better because of two main reasons. First, larger tile sizes lead to
more parallel hardware as the HLS can unroll larger innermost
intra-tile loops. Second, the unrolled basic blocks lead to
bigger/deeper pipelines that can better hide memory latencies.
Theoretically this can explain a factor of 2$\times$. The range we see in the
numbers is caused by both the number/sizes of the cache buffers and
the amount of reuse. The data must be loaded into the buffers before
the tiles access it. For full buffers the HLS cannot overlap the
filling of the buffers and the execution of a tile, as the data needs
to be in the buffer before it is accessed. For chunk buffers we
carefully crafted the pre-loading of segments of the data so that the
HLS can partly overlap the loading with \ship{} operations further
down the tile's loop nest. For full buffers and, because of the
pre-loading, even for chunk buffers, each load has a fixed runtime
overhead. The more often the tile uses a data in the buffer, the
better it can amortize these fixed costs. This explains why we see
factors around 2x for the high-point stencils.  The low-point stencils
have factors around 1.3x.  (\texttt{2D-9p} and \texttt{3D-heat} are
mid-tier.)  The exception to the rule are the \texttt{fdtd}
stencils. Despite being 4-point stencils they strongly benefit from
larger tile sizes. The reason is their in-place updates. Whereas our
optimized FPGAs can map them to fast register operations, the baseline
FPGA always keeps them sequential.

\begin{figure}
\pgfplotsset{width=\linewidth, height=4cm}
\centering
\begin{tikzpicture}
\begin{axis}[
    ybar=1.44pt,
    bar width=5pt,
    ylabel={speedup},
    symbolic x coords={1D-jacobi,2D-5p,2D-9p,2D-jacobi,2D-fdtd0,2D-fdtd1,2D-fdtd2,3D-19p,3D-27p,3D-heat},
    xtick=data,
    x tick label style={rotate=40, anchor=east, inner sep=0pt},
]

\addplot[pattern=horizontal lines gray]
coordinates {(1D-jacobi, 70.5197063)(2D-5p, 39.41469045)
(2D-9p, 76.21239902)(2D-jacobi, 0)(2D-fdtd0, 118.5917425)(2D-fdtd1,
129.035354)(2D-fdtd2, 55.20542621)(3D-19p, 61.93852411)(3D-27p,
77.98415675)(3D-heat, 29.38813632)};

\addplot[pattern=north east lines] coordinates {(1D-jacobi, 109.7340615)(2D-5p,
70.06574608)(2D-9p, 134.1088598)(2D-jacobi, 86.26127252)(2D-fdtd0,
156.9009508)(2D-fdtd1, 0)(2D-fdtd2, 76.33073038)(3D-19p,
96.83004876)(3D-27p, 117.7512451)(3D-heat, 43.90319504)}
node[pos=0.4, anchor=west, xshift=.8ex] {\small156}
node[pos=0.95, anchor=west, xshift=.8ex] {\small43}
;

\end{axis}

\end{tikzpicture}
\caption{
  Speedups
  of two optimized FPGAs for port widths
  8 \protect\tikz\protect\node[draw, rectangle, pattern=horizontal
  lines gray, minimum width=1em]{}; and 16
  \protect\tikz\protect\node[minimum width=1em, draw, rectangle,
  pattern=north east lines]{};
  (and large
  tile sizes $SZ_{1D}$=$(1024)$, $SZ_{2D}$=$(128,128)$, $SZ_{3D}$=$(32,32,32)$).
    Empty bar: as before.
}\label{fig:ports}
\end{figure}

\paragraph{Impact of the port width.}
\fig{ports} shows that wider bursts boost performance as they increase
memory throughput and fill buffers faster. Note, that the second bar of
each benchmark is the same as in \fig{tiles}. As mentioned before, even though
HLS can sometimes partly overlap the filling of the buffers with the
unrolled/pipelined tile, there are still fixed costs. Hence,
a quicker buffer initialization has a direct impact on
the overall runtime performance. The smaller the runtime of a tile is,
the harder it is for the tile to amortize the fixed costs,
and hence the more impact an increased port width has. The impact of
the port width is stronger whenever the impact of the tile size is
smaller, and the other way round. This is exactly what a comparison of
\fig{tiles} and \fig{ports} reveals.\\

We do not show bars for the combination of two bad choices, i.e., for
small tile sizes and a small burst width, since this leads to an even
stronger drop of the achieved speedups.

\paragraph{Importance of the right permutation and of padded tile loops.}
Let us now demonstrate that some of the choices in \secref{impl} are
crucial for the performance. Instead of working with all benchmark
codes we pick \texttt{2D-5p} as it is affected the least by bad
choices.  For the other benchmarks the effect of a wrong choice is
even stronger.  We do not consider the three \texttt{fdtd} codes
because of their irregular in-place behavior. And we also ignore the
stencils for which the HLS refuses to generate an FPGA.  The
5-point stencil \texttt{2D-5p} is computationally the most simple one
in the remaining set.

In \secref{impl}, step 2 picks the permutation that needs the smallest
total buffer size. For the 10 stencils, it considers 1--6
potential permutations. For each benchmark there are two
groups of permutations. The more costly group requires 2 or 3 full
buffers whereas the cheaper group can do without any full buffers.
For the total runtime of a stencil on the FPGA, we need to
look at the sum of the buffer filling time and the tile's execution
time, it is the number and the size of the full buffers that make the
selection of a wrong permutation most obvious. Compared to 3D
stencils, 2D stencils like \texttt{2D-5p} are less affected by a
wrong choice of buffer type as their full buffers are smaller (as
they only need to store 2D data).

For \texttt{2D-5p} we saw in \fig{tiles} a speedup of 70$\times$ over the
baseline. For this speedup, step 2 picked a chunk buffer and a line
buffer. When picking a wrong permutation instead, we see a smaller
speedup. The achieved speedup of the generated FPGA drops
from 70$\times$ down to 21$\times$. Picking the best permutation is even
more important for computationally more demanding stencils with larger
working sets.

Step 3 of \secref{impl} removed the \texttt{min}-operation from
the innermost loop. This caused the last tile to always be a full
one. The rationale that this enables the HLS to
unroll/pipeline the innermost loop is justified in the 10
benchmarks. For code versions that still have the
\texttt{min}-operations the HLS generates sequential hardware for the
innermost loop; the bodies are no longer executed in parallel. (The
HLS protocols reveal that there is no unrolling and the consumption
of DSP blocks shrinks by 8$\times$--22$\times$
which indicates sequential behavior.)

When we switch off the padding of the innermost tile loop of
\texttt{2D-5p}, there is not just a reduced speedup. The resulting
FPGA is even \textit{slower} than the baseline FPGA by a factor of
2$\times$ (whereas the padded version saw a speedup of 70$\times$).
Loop padding is even more important for computationally more demanding
stencils.

\section{Conclusion}\label{sec:fw}
This paper
presents an optimization that
offloads memory-bound stencil codes to FPGAs that are
$43\times$ and $156\times$ faster than the hardware that the HLS
generates with its default optimization settings.  The presented
technique uses polyhedral methods to pick the best loop tiling that --
when inlined cache buffers are added to exploit data-reuse -- achieves
the best runtime.  Other key insights are that the  iteration space
of the intra-tile loops should be padded to avoid incomplete tiles, that
heavy tiling of the iteration space enables the HLS to unroll/pipeline
the tiles into concurrent hardware, and that by adding halo elements to
the cache buffers and properly aligning the memcopy operations, the
HLS can utilize wide bursts to further boost the runtime performance.

\bibliographystyle{ACM-Reference-Format}
\bibliography{impact}

\end{document}